\documentclass[onecolumn,authoryear]{els-mrw} 

\usepackage{amsmath,amssymb,amsfonts,amsthm,makeidx,graphicx}
\usepackage{txfonts}
\usepackage{helvet}

\begin{document}

\chapter{Giants, Supergiants and Hypergiants}\label{chap1}

\author[1]{Lee R. Patrick}%


\address[1]{\orgname{Centro de Astrobiolog\'ia}, \orgdiv{(CSIC-INTA)}, \orgaddress{Ctra.\ Torrej\'on a Ajalvir km 4, 28850 Torrej\'on de Ardoz, Spain}}

\articletag{Chapter Article tagline: update of previous edition,, reprint..}

\maketitle

\begin{abstract}[Abstract]
Observing the stars in our night sky tells us that giant, supergiant and hypergiant stars hold an unique importance in the understanding of stellar populations. 
Theoretical stellar models predict a rich tapestry of evolved stars. 
These evolved stars result in supernova explosions for massive stars and the shedding of the outer layers of low- and intermediate-mass stars to form  white dwarf stars. 
Stars on these pathways provide elemental and kinematic feedback that shapes their host galaxies and synthesize the elements that we observe in the Universe today. 
Observational surveys in the Milky Way and the Local Group of galaxies shape our physical understanding of stellar evolution. 
Hypergiant stars represent stellar evolution at the extremes, displaying evidence of intense stellar winds combined with luminosities and radii that are unrivalled.
Supergiant stars are used as elemental beacons in distant galaxies, and supergiants that cross the Cepheid variable instability strip can be used as rulers to measure the scale of our Universe. 
The overwhelming majority of stars in the Milky Way will become giant stars.
Giants are hundreds of times larger and thousands of times more luminous than the Sun, which makes them excellent tracers of stellar structure. 
In this chapter, I describe the observational and theoretical manifestations of stellar evolution at various stages and masses. I highlight evolutionary connections between evolved products and comment on their eventual endpoints.
\end{abstract}

\begin{glossary}[Glossary]

\term{Blue loop} A feature of stellar evolutionary models in which a helium-burning giant or supergiant evolves blueward to spend a fraction of its helium-burning lifetime at warmer temperatures. 

\term{Cepheid variable} A pulsating star that is located in the instability strip of the HRD. Cepheid variables are giant and supergiant stars of spectral type F to K. 

\term{Dwarf} A star currently fusing hydrogen to helium in its core. Most stars spend 90\,\% of their lives as dwarfs. See also \textbf{Giant}, \textbf{Supergiant} and \textbf{Hypergiant}.

\term{Giant} A star more luminous than a dwarf of the same spectral type. Giants are typically evolved stars. See also \textbf{Dwarf}, \textbf{Supergiant} and \textbf{Hypergiant}.

\term{Hypergiant} A star of extreme luminosity that displays features of strong mass loss. See also \textbf{Dwarf}, \textbf{Giant} and \textbf{Supergiant}.

\term{Local Group} The Local Group of galaxies is a group of at least 50 galaxies that contain the Milky Way. The Local Group consists of two large galaxies, the Milky Way and M31 and various satellite galaxies. The LMC and SMC are satellite galaxies to the Milky Way. See Also \textbf{Magellanic Clouds} and \textbf{Milky Way}.

\term{Luminosity class} A classification scheme to divide stars into groups with similar luminosities. This ranges from dwarf (V) to giant (III) to hypergiant (Ia$^+$).  See also \textbf{Dwarf}, \textbf{Giant}, \textbf{Supergiant} and \textbf{Hypergiant}.

\term{Luminous blue variable} A rare supergiant or hypergiant star that undergoes eruption events. 

\term{Magellanic Clouds} The LMC and SMC are two dwarf galaxies that are located near the Milky Way. These stellar systems represent two of the best-studied stellar populations outside the Milky Way. While significantly less massive than the Milky Way, these galaxies have a rich population of giant, supergiant and hypergiant stars. See also \textbf{Milky Way}.

\term{Massive star} A star that is born with sufficient mass to explode as a supernova. 

\term{Milky Way} The Milky Way is a barred spiral galaxy that hosts on the order of 200 billion stars. The Solar system is in the Milky Way. 

\term{Supergiant} A star more luminous than a giant of the same spectral type. Supergiants are almost exclusively evolved stars.  See also \textbf{Dwarf}, \textbf{Giant} and \textbf{Hypergiant}.

\term{Stellar winds} An outflow from a star by means of which mass is lost.

\term{Supernova} A large stellar explosion. Rarer examples occur, but supernovae come in two broad types: thermonuclear and core-collapse. A core-collapse supernova is the endpoint of the evolution of a massive star. RSGs are the progenitors of the most common type of core-collapse supernova (Type II-P). BSGs and YSGs have also been identified as progenitors of core-collapse supernova. The progenitor of SN1987A in the LMC was famously a BSG star, rather than a RSG as had been predicted. In rarer cases, BHGs and YHGs have also been identified as core-collapse supernova progenitors, but this is not universally accepted in the literature. 

\term{Terminal-age main sequence (TAMS)} The point at which a star ceases hydrogen fusion in the core. See also \textbf{Zero-age main sequence}.

\term{Zero-age main sequence (ZAMS)} The point at which a star begins hydrogen fusion in the core. See also \textbf{Terminal-Age main sequence}.
\end{glossary}

\begin{glossary}[Nomenclature]
\begin{tabular}{@{}lp{34pc}@{}}
AGB & Asymptotic giant branch\\
BHG &Blue hypergiant star\\
BSG &Blue supergiant star\\
IMF & Initial mass function\\
HRD & Hertzprung--Russell diagram\\
LBV & Luminous blue variable star\\
LMC & Large Magellanic Cloud\\
OB star & Star of spectral type O or B\\
RGB & Red giant branch\\
RHG &Red hypergiant star\\
RSG &Red supergiant star\\
SMC & Small Magellanic Cloud\\
TAMS &Terminal-age main sequence\\
YSG &Yellow supergiant star\\
YHG &Yellow hypergiant star\\
WNEh & Early nitrogen-rich Wolf--Rayet star\\
WNh & Nitrogen-rich Wolf--Rayet star\\
WNLh & Late nitrogen-rich Wolf--Rayet star \\
ZAMS & Zero-age main sequence\\
\end{tabular}
\end{glossary}

Key points
\begin{itemize}
    \item Giant, supergiant and hypergiant stars are luminous, evolved, stars that represent some of the brightest and largest stars in star-forming galaxies. 
    \item While giant, supergiant and hypergiant stars are generally a late stage of stellar evolution where core hydrogen fuel has been exhausted, O and early-B type stars spend a fraction of their main-sequence lifetime as giant, supergiant and even hypergiant stars. This is a function of mass and stars above 60\,M$_\odot$ are supergiant and hypergiant stars even on the very early main sequence. As OB stars age they evolve through the luminosity classes, evolving from giants, to supergiants to hypergiants.
    \item Hypergiant stars are brief phases in the lifetimes of massive stars. BHGs, YHGs and RHGs are highly variable and hypergiants are known to undergo extreme outbursts related to the LBV phenomenon. An evolutionary connection is likely to exist between RHGs and YHGs stars which may further extend to late-BHGs.
    \item The terminology hypergiant is reserved exclusively for massive stars, whereas giant and supergiant are terms used for low, intermediate and massive stars. 
    \item Binary interaction products are likely to be a key evolutionary pathway that must be invoked to explain the statistics of BSGs, YSGs and RSGs.
    \item RSGs are the evolutionary phase immediately proceeding core-collapse supernova for the majority of massive stars. RHGs are high-luminosity examples of RSGs that exhibit extreme mass-loss rates. 
    \item Low- and intermediate-mass stars evolve onto the RGB after core-hydrogen burning. These stars evolve through the AGB phase before shedding their outer envelopes to become white dwarf stars. 
\end{itemize}

\section{Introduction}\label{intro}

Giant, supergiant and hypergiant are examples of luminosity classes of stars in the Morgan \& Keenan classification system~\citep[][]{Morgan1973}. These terms are used to describe stars that appear \textit{spectroscopically} more luminous than dwarf stars. 
This observational classification scheme was later found to have underlying physical meaning in that dwarf stars, which are
are assigned a Roman numeral V, are currently fusing hydrogen to helium in their cores. This represents around 90\,\% of the lifetime of stars. 

Stars classified as giants and supergiants are assigned Roman numerals from III to I, with this numeral representing a luminosity progression:
supergiants (I) are more luminous than giants (III). 
\citet{1942ApJ....95..461K} split luminosity class I into lower and higher luminosities and classified the red supergiant (RSG) RW~Cep as luminosity class 0, singling this star out -- along with VV~Cep, which is now known to be a binary system~\citep{1977JRASC..71..152W} -- as too luminous to correctly fit into the supergiant luminosity class.
Because of this, a hypergiant luminosity classification was introduced, first as super-supergiants by \citet{1956MNRAS.116..587F}, then latterly as hypergiants by~\cite{1980GAM....19.....D}. 
Hypergiants are assigned a luminosity class 0 or -- more commonly in the literature -- a luminosity class Ia$^{+}$.
For the remainder of this chapter, the more common Ia$^{+}$ is used. 
In addition, the supergiant luminosity class is further divided into three subclasses: Ia, Iab and Ib.
This, again, represents a luminosity progression, where stars with the Ia classification are the most luminous of their class.
For more details on the classification of stars, see the \textit{Stellar Classification} chapter.

\begin{figure}[t]
\centering
\includegraphics[width=.9\textwidth]{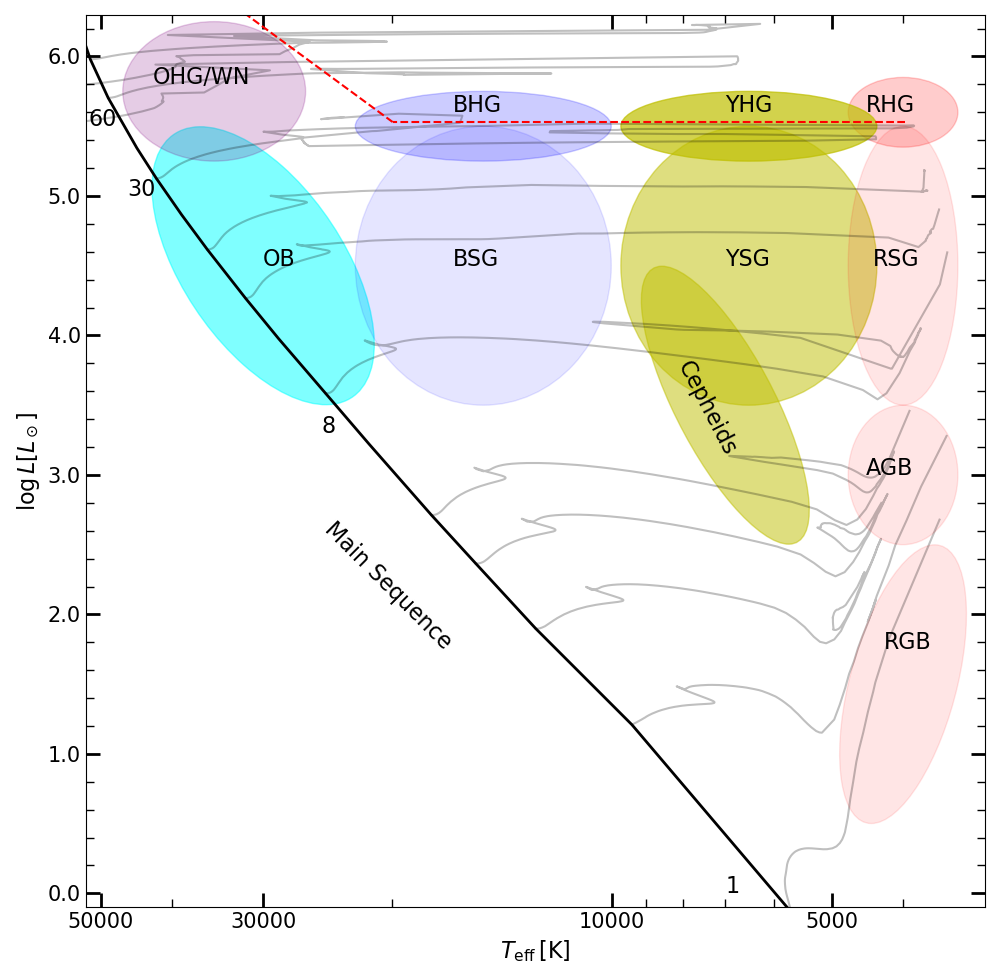}
\caption{Hertzprung--Russell Diagram (HRD) with stellar evolutionary models ranging from 1 to 85\,M$_\odot$. This figure shows the ranges in stellar parameters that are covered by the different groupings of giants, supergiants and hypergiants in this chapter. Evolutionary models are non-rotating tracks from~\citet{2012A&A...537A.146E}. The zero-age main sequence (ZAMS) is shown with a solid black line by interpolating between the initial points of the stellar models. The empirically defined upper limit on stellar luminosities, the Humphreys--Davidson (HD) limit, is shown with a red dashed line.}
\label{fig:HRD}
\end{figure}

The best way to visualise stellar evolution is on the so-called Hertzsprung--Russel diagram (HRD). 
This is displayed in Figure~\ref{fig:HRD}, where the realms of the giants, supergiants and hypergiants are marked.
A HRD shows stellar effective temperature on the ordinate against luminosity on the abscissa.
Luminosity and effective temperature are two fundamental observable quantities that allow us to understand the physical processes that determine stellar evolution. 
When their luminosities and temperatures are shown on the HRD, dwarf stars appear on an observational sequence termed the main-sequence of stellar evolution, which is illustrated in Figure~\ref{fig:HRD}.
The grey tracks on this figure represent theoretical evolutionary sequences of stars for masses in the range between 1 and 85\,M$_\odot$. 
Marked with different coloured shapes are the location of the giants, supergiants and hypergiants that are discussed in this chapter.
Viewing these various classifications on the HRD highlights the fact that the process of dividing a continuous sequence of stars in a small number of luminosity bins based on purely observational characteristics results in overlapping and often arbitrary distinctions between different classes.
This also results in very similar terms being used for very different stars.
The red giant and red supergiant distinction is an example of this:
these stars are often very similar in appearance but represent very different stages of stellar evolution. 

Because of its broad nature, this chapter encompasses various types of stars, which have chapters of their own in this encyclopedia. 
Despite later stages of stellar evolution accounting for only around 10\,\% of the lifetime of stars, giants, supergiants and hypergiants are vital tools that are used to test our understanding of stellar astrophysics. 
Because of their importance, the literature on giants, supergiants and hypergiants is vast. 
Therefore, where appropriate at several points in this chapter, the reader is referred to other chapters in this book, as well as more in-depth reviews. 
The aim of this chapter is to provide an overview of evolved stellar products. To this end, some basic theoretical and observational concepts are outlined in Section~\ref{basics}. 
Sections~\ref{ob}, \ref{ysgs} and~\ref{rsgs} roughly follow the lives of massive stars as they evolve from the main sequence, through short-lived evolutionary phases as blue, yellow and red supergiants. 
Section~\ref{giants} describes the giants and supergiants that have evolved from low- and intermediate-mass stars as they move through the red giant branch (RGB).
Finally, Section~\ref{summary} presents some key summary points.






\section{Underlying theoretical and observational concepts} \label{basics}
In this section, some basic theoretical and observational principles are described that will aid in the understanding of the concepts and evolutionary connections encountered in the later sections of this chapter. 

\textbf{Initial mass function:}
The distribution of stars as a function of mass at birth is described as the the \textit{initial mass function} (IMF).
Such a function can take various forms and whether the IMF is a constant throughout the Universe is an open question~\citep[see the review by][]{2010ARA&A..48..339B}.
The most commonly used functional form of the IMF is arguably that of~\citet{1955ApJ...121..161S}, which is $\chi (m) \propto m^{-\alpha}$, where $\alpha=2.35$.
The general form of this function tells us that forming more massive stars is more difficult in nature and, therefore, high-mass stars are intrinsically rarer.
A simple example using the Salpeter IMF shows that for every 10\,M$_\odot$ star that is born there are 224 1\,M$_\odot$ stars.

\textbf{Eddington limit:}
The Eddington limit is the critical luminosity above which a star can no longer balance the radiative force resulting from the energy production in the stellar interior.
The Eddington luminosity for hot stars, where electron scattering is the dominant source of opacity, is given by the equation,
\begin{equation}
    L_{\rm Edd} = \frac{4\pi G M m_{\rm p} c}{\sigma_{\rm T}},
\end{equation}
where $L_{\rm Edd}$ is the Eddington luminosity, $G$ is the gravitational constant, $M$ is the mass of the star, $m_p$ is the mass of a proton, $c$ is the speed of light and $\sigma_{\rm T}$ is the Thomson scattering cross-section for an electron.
The Eddington luminosity may be reached for hot massive stars during core-hydrogen burning. 
In addition, luminous blue variables (LBVs) are located near the Eddington limit. 

\textbf{Hayashi limit:}
The Hayashi limit is a theoretical limit on the radius of a star within hydrostatic equilibrium.
The Hayashi limit~\citep{1961PASJ...13..442H} is a boundary beyond which a star is no longer stable. This is important for evolved stars that have expanded after core-collapse. Such stars reach the Hayashi limit, which in effect limits their radii and effective temperatures. 
This limit is a function of mass and chemical composition of the star. 
In the HRD, the Hayashi tracks are almost a vertical limit at around an effective temperature of 4000\,K.

\textbf{Humphreys--Davidson limit:}
The Humphreys--Davidson (HD) limit is the result that cool luminous stars are not observed above a threshold luminosity~\citep{1979ApJ...232..409H}.
This limit is illustrated on the HRD in Figure~\ref{fig:HRD} and is observed in the Milky Way and Local Group galaxies.

\begin{figure}[t]
\centering
\includegraphics[width=.5\textwidth]{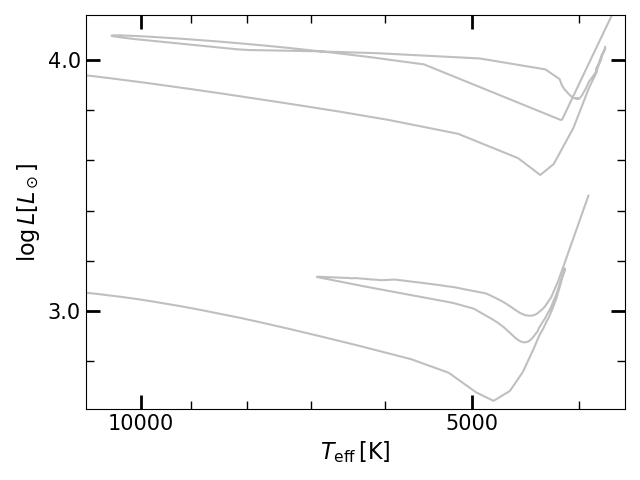}
\caption{A Hertzsprung--Russel diagram (HRD) for 5 and 9\,M$_\odot$ stars, highlighting the morphology and extent of the blue-loop feature.}
\label{fig:blueloop}
\end{figure}

\textbf{Blue loops:}
During core-helium burning, for stars in the rough mass range 3 to 10\,M$_\odot$, some stellar models predicted a phenomenon called a `blue loop'. 
Stellar evolutionary models predict that once a star evolves off the main sequence, it rapidly evolves onto the red giant phase on a thermal timescale. 
A blue loop is where red giant and supergiant stars undergo a period of evolution from red to blue on a nuclear timescale.
Figure~\ref{fig:blueloop} illustrates the blue loop phenomenon. 
Blue loops occur when a star is in an almost vertical evolutionary phase determined by the Hayashi limit. 
This phenomenon is triggered by the outward evolution of hydrogen shell burning that arrives at a layer of excess helium.
The effect of blue loops is that the star appears as an A- to K-type supergiant star for a period of time during its core-helium burning phase. 
The mass of stars that undergo blue loops and the extension to the blue are dependent on metallicity. 
\citet{2015MNRAS.447.2951W} provide a more in-depth description of blue loops. 
 
Some Cepheid variables are expected to be experiencing a blue loop. More generally, blue-loop evolution is one of the potential explanations for the observed overabundance of blue supergiants (BSGs) in Local Group galaxies.

\section{OB stars: Giants, supergiants and hypergiants on the main sequence} \label{ob}


Massive stars fusing hydrogen to helium on the main sequence appear observationally as O- and B-type stars depending on their mass. 
During this time, their lives are dominated by intense radiation-driven stellar winds that are millions of times greater than those experienced for solar-like stars (see the \textit{Stellar Winds} and \textit{Stellar Atmospheres} chapters).
Such a high intrinsic rate of mass-loss affects their evolution considerably, causing envelope self-stripping and resulting in the stars appearing as main-sequence Wolf--Rayet stars at the highest masses~\citep[][and see the \textit{Wolf--Rayet stars} chapter]{1997ApJ...477..792D}. 
As sources of ionising radiation in the early Universe, OB stars provided the radiation that played an essential role in the epoch of reionisation of the Universe~\citep{1997ApJ...483...21H}. 
In addition, the intense stellar winds act to distribute the products of nucleosynthesis throughout their host galaxies and, in this sense, massive stars are the drivers of the elemental and kinematic evolution of galaxies on large scales.
In the OB star region, because of the almost vertical shape of the main sequence in the HRD, OB stars do not spend all of their main-sequence lifetimes as dwarf stars. 


One of the key complicating factors in the lives of massive stars is multiplicity~\citep{Langer2012}. The majority of massive stars are born within a close binary system, such that around 70\,\% are expected to interact with a companion within their lifetimes~\citep[e.g.][]{2014ApJ...782....7D}. 
As binaries interact, they drastically alter their appearances and a rich population of interaction products is observed.
More evolved giants, supergiants and hypergiants are all affected dramatically by binary evolution and, in this respect, massive main-sequence stars represent a key population in which to test stellar physics before binary interaction takes places. 
This is vital to
set the foundations of the physical properties that then go on to predict further evolution.

\subsection{Stellar structure and evolution}
OB stars fuse hydrogen to helium in their cores on the main sequence via the highly temperature dependent CNO cycle.
As such, a convective core is established, which allows fusion products to be mixed into the upper atmospheric layers. 
As core-hydrogen burning proceeds, the star increases in luminosity and decreases in temperature as a response to the star expanding. 
This evolution causes the star to evolve from a dwarf to a giant to a supergiant, all while on the main sequence.
This evolution is unique to O- and early-B-type stars: at all other spectral types, giants, supergiants and hypergiants are effectively unrelated to dwarf stars of the same spectral type.

\begin{figure}[t]
\centering
\includegraphics[width=.5\textwidth]{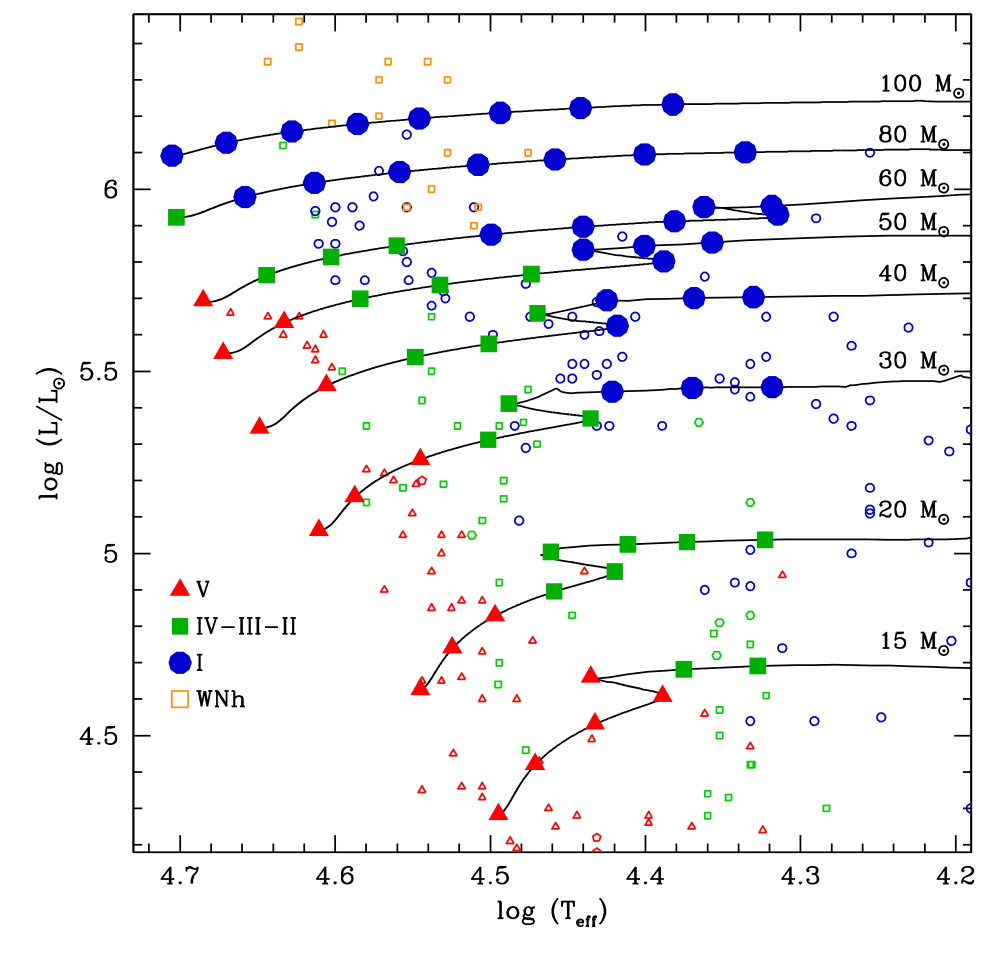}
\caption{Hertzprung--Russell Diagram (HRD) of massive OB-type stars showing the progression in luminosity classification of spectra extracted from stellar models for masses in the range 15 to 100\,M$_\odot$ (large filled symbols), alongside observed classifications of massive stars (smaller non-filled symbols). An example of the location of the end of the main sequence in the stellar tracks is labelled. One can see that only for the 15\,M$_\odot$ stellar track do the stars appear as a luminosity class V star (i.e. a dwarf on the main sequence) for the entire main sequence. Adapted from~\citet{2017A&A...598A..56M}. }
\label{fig:mp17-hrd}
\end{figure}

More massive stars appear as earlier spectral types.
An O-type dwarf evolves into an O-type giant, of later spectral subclass, which then subsequently evolves into a supergiant -- again of later spectral class. 
Figure~\ref{fig:mp17-hrd} shows a HRD illustrating this effect.
This figure shows spectral sequences from simulated observations extracted from stellar models, which were then classified as if they were observations.
Various authors have refined such a calibration between spectral classification and intrinsic stellar properties~\citep[mass, luminosity, effective temperature;][]{1997A&A...322..598S,2010A&A...524A..98W,2014A&A...564A..30G,2017A&A...598A..56M}. 
At higher masses, O-type stars do not appear as dwarf stars; instead, they appear as giant or supergiant stars, even on the ZAMS~\citep{2014A&A...564A..30G,2017A&A...598A..56M}. 
The following schematic is an illustrative guide on evolution within the main sequence for OB-type stars:

\begin{itemize}
    \item[] \, 8--16\,M$_\odot$: B~V $\rightarrow$ B~III
    \item[] 16--30\,M$_\odot$: O~V $\rightarrow$ O~III $\rightarrow$ BI 
    \item[] 30--60\,M$_\odot$: O~III $\rightarrow$ O~I $\rightarrow$ B~I$^{+}$
    \item[] 60--80\,M$_\odot$: O~I $\rightarrow$ O~I$^{+}$ $\rightarrow$ WNLh
    \item[] \,~$>80$\,M$_\odot$: WNEh $\rightarrow$ WNLh

\end{itemize}

This guide is compiled from multiple observational studies, including~\citet{Martins2008},
~\citet{2012A&A...541A.145C,clark2018A,2019A&A...623A..84C} and \citet{2016MNRAS.458..624C}. 
These sequences should be considered illustrative and the mass ranges -- particularly the higher masses -- are approximate. 
\citet{Langer2012} provides a graphical view of such evolutionary sequences, which nicely illustrates the length of time spent in different evolutionary phases.
As noted above, these evolutionary connections can be significantly altered by the effects of binary evolution
(see the chapters on \textit{Observing Binary Stars}, \textit{Multiple Star Statistics} and \textit{Evolution of Binary Stars} for further details).

The evolution and eventual fate of the highest mass stars -- which are often rather loosely termed \textit{very massive stars}\footnote{\citet{2015ASSL..412.....V} reviews \textit{very massive stars} and offers a definition of above 100\,M$_\odot$, but this is not universally adopted in the literature and this term can indicate stars above around 60\,M$_\odot$} -- are uniquely important in the understanding of galaxies in the early Universe~\citep[see][]{2022ARA&A..60..455E}.
However, partly as a result of their intrinsic rarity, evolutionary pathways have been difficult to establish~\citep[e.g.][]{1994A&A...290..819L,1995A&A...293..427C,2006A&A...457.1015H}.
For stars with initial masses above $\sim$80\,M$_\odot$, empirical evidence shows that evolution transitions from early-O-type supergiant to O-hypergiant.
As with hypergiant stars of later spectral type, these stars are differentiated based on evidence of extreme mass-loss rates manifesting as intense emission and P~Cygni line profiles in their spectra~\citep{2012A&A...541A.145C,2015ApJ...809..109W}.
Beyond this, the nitrogen-rich Wolf–Rayet (WNh) stars represent an even more extreme evolutionary phase of stars of masses greater than around $80-100$\,M$_\odot$~\citep[e.g.][]{Tramper2016,clark2018A,Lohr2018,2023MNRAS.521.4473C}. 
Based on observations in young massive clusters in the Milky Way, a tight evolutionary connection is revealed between supergiants, hypergiants and WNh stars~\citep{clark2018A}, which may subsequently evolve into LBVs~\citep{2008ApJ...679.1467S} and back again~\citep{clark2018Q}.
WNh stars are described in more detail in the \textit{Wolf--Rayet Stars} chapter.


\subsection{Observed populations} 

Despite the progress made in delineating evolutionary sequences for massive stars within the last 30 years, between the various O-type spectral classifications an intrinsic problem remains that an individual spectral classification -- at a given metallicity -- corresponds to a large range of masses for stars at different evolutionary stages~\citep[see][for an illustration on the HRD]{2010A&A...524A..98W}.
To break this degeneracy requires knowledge of the age of the star.
In this respect, massive stellar clusters are the only systems where the ages of entire populations of massive stars are effectively fixed and relatively well understood.
Such young massive clusters are vital in our understanding of massive stars in general and in particular the evolutionary connections in the OB star regime.
\citet{2010ARA&A..48..431P} review the properties of young massive star clusters in the Milky Way and Local Group of galaxies. 

In the central regions of the Milky Way, with near-infrared (IR) observational techniques multiple massive star clusters were identified in the 1990s~\citep{1996ApJ...461..750C,1999ApJ...525..750F}. 
The massive star population in these clusters and that in an apparently isolated population separate from the main clusters have been shown to be uniquely rich~\citep{2010ApJ...725..188M,2021A&A...649A..43C}.
Within this context, the Arches, Quintuplet and Galactic Centre clusters are important populations.
Because of limitations relating to the extreme, variable extinction in this region, spectroscopic surveys have focused mainly on the brightest giants, supergiants and hypergiants.
These studies have shown that the clusters populate the upper reaches of the IMF, which is intrinsically rare and makes these clusters highly valuable to study coeval populations of high-mass stars. 
Outside the central regions of the Milky Way, the Westerlund-1 cluster has for many decades been a reference point for massive star evolution~\citep[see][and references therein]{2022A&A...664A.146N}, and this cluster also hosts a unique cohort of evolved supergiants and hypergiants~\citep{2010A&A...516A..78N,2010A&A...520A..48R,2024ApJ...964..171B}. 
Outside of the Milky Way, the R136 cluster hosts some of the most massive stars known in the Local Universe~\citep{2016MNRAS.458..624C}. 
Within the Large Magellanic Cloud (LMC), this cluster is located within the Tarantula nebula, which hosts a rich population of slightly older massive stars~\citep{2006A&A...456..623E,2011A&A...530A.108E}.

Large-scale efforts are underway to catalogue the Milky Way's population of O-type stars. 
Surveys such as GOSS~\citep{2013msao.confE.198M} have an online database of O-type stars.
The IACOB~\citep{2014A&A...562A.135S} and the OWN surveys~\citep{2017IAUS..329...89B} focus on detailed high-resolution spectroscopic observations of OB-type stars at various evolutionary stages.
Extragalactic studies of massive stars allow important metallicity effects to be examined and, in particular, low-metallicity environments (Small Magellanic Cloud (SMC) metal content and lower) provide a connection to stars with the same elemental abundances as those in the peak of the star formation in the Universe~\citep{Madau2014}.
\citet{2013NewAR..57...14M} reviewed the massive star populations in the star-forming galaxies of the Local Group and highlighted photometric and spectroscopic surveys that have provided advances in our understanding.
The low-metallicity environment of the SMC has represented something of a frontier in stellar evolution for decades; however, progress is starting to be made on multiple fronts.
\citet{2022MNRAS.516.4164L} identified over 150 OB-type stars in the Sextans~A galaxy, which has an oxygen abundance of 1/15--1/10 that of the solar value~\citep{Skillman1989} and around 1/10 of the iron solar abundance~\citep{Kaufer2004}.
In addition, O-type stars have been identified in the bridge between the LMC and SMC galaxies~\citep{2021A&A...646A..16R}, a fraction of which appear to have significantly lower metal content than the SMC.

A list of the known WNh stars together with other Wolf--Rayet stars was published by \citep{2015MNRAS.447.2322R}.\footnote{An up-to-date list is maintained at the following link: \url{http://pacrowther.staff.shef.ac.uk/WRcat/index.php}} 
A corresponding sample of O-type hypergiants is, however, not available in the literature or in the online GOSS database. 
Between the Arches, Quintuplet and Galactic Centre clusters and the isolated massive star population, the stellar population at the Galactic Centre hosts a rich collection of O-type hypergiant and supergiant stars (see~\citet{2021A&A...649A..43C} and references therein for more details on the observational properties of these samples).

\section{Blue and yellow supergiants and hypergiants}\label{ysgs}
Blue and yellow supergiants (YSGs) and hypergiants are post-main-sequence stars that are in general believed to be in a brief transitionary phase evolving towards the red. 
At the blue border, 
the division between main-sequence stars discussed in the previous chapter and some of the post-main-sequence stars discussed here is not clear observationally or well understood theoretically.
While supergiants and even hypergiants in the O-type star domain can be reasonably safely considered main-sequence stars, for early B-type supergiants and hypergiants the situation is less clear~\citep{BLOeM-BI}.
Similarly, on the red border, because of their relative rarity, the distinction between yellow and red supergiants is often not clear observationally. 
As can be seen in Figure~\ref{fig:HRD}, the evolutionary pathways of hypergiant stars is more complicated than a simple evolutionary sequence from blue to red and, while not illustrated in Figure~\ref{fig:HRD}, this is also true to some extent for the supergiants. 
The observed statistics of BSGs and YSGs is a sensitive test of evolutionary models, which can calibrate observationally the terminal-age main sequence (TAMS).
In addition, the appearance of blue loops and evolutionary connections with RSGs are outstanding issues that implicate BSGs and YSGs~\citep{BLOeM-BAF}, as well as their more extreme hypergiant counterparts. 



\subsection{Hypergiants}
Hypergiant stars are a class of objects that show evidence of extreme luminosities and mass-loss rates. 
Because of their high masses, combined with the brevity of the different evolutionary phases, hypergiant stars are rare, and only handfuls of examples are known in the Milky Way and Local Group galaxies.
Blue hypergiants (BHGs) and yellow hypergiants (YHGs) have spectral types in the B--G range.
Because of a rough evolutionary grouping, O-type hypergiant stars are considered separately in Section~\ref{ob} and red hypergiants (RHGs) are considered with the RSGs in Section~\ref{rsgs}.
In general, similar to the less luminous BSGs, BHGs and YHGs are an inhomogeneous group of objects that consist of stars from different evolutionary pathways that broadly occupy the same physical properties.

\begin{figure}[t]
\centering
\includegraphics[trim=0cm 3cm 0cm 3cm, clip,width=.95\textwidth]{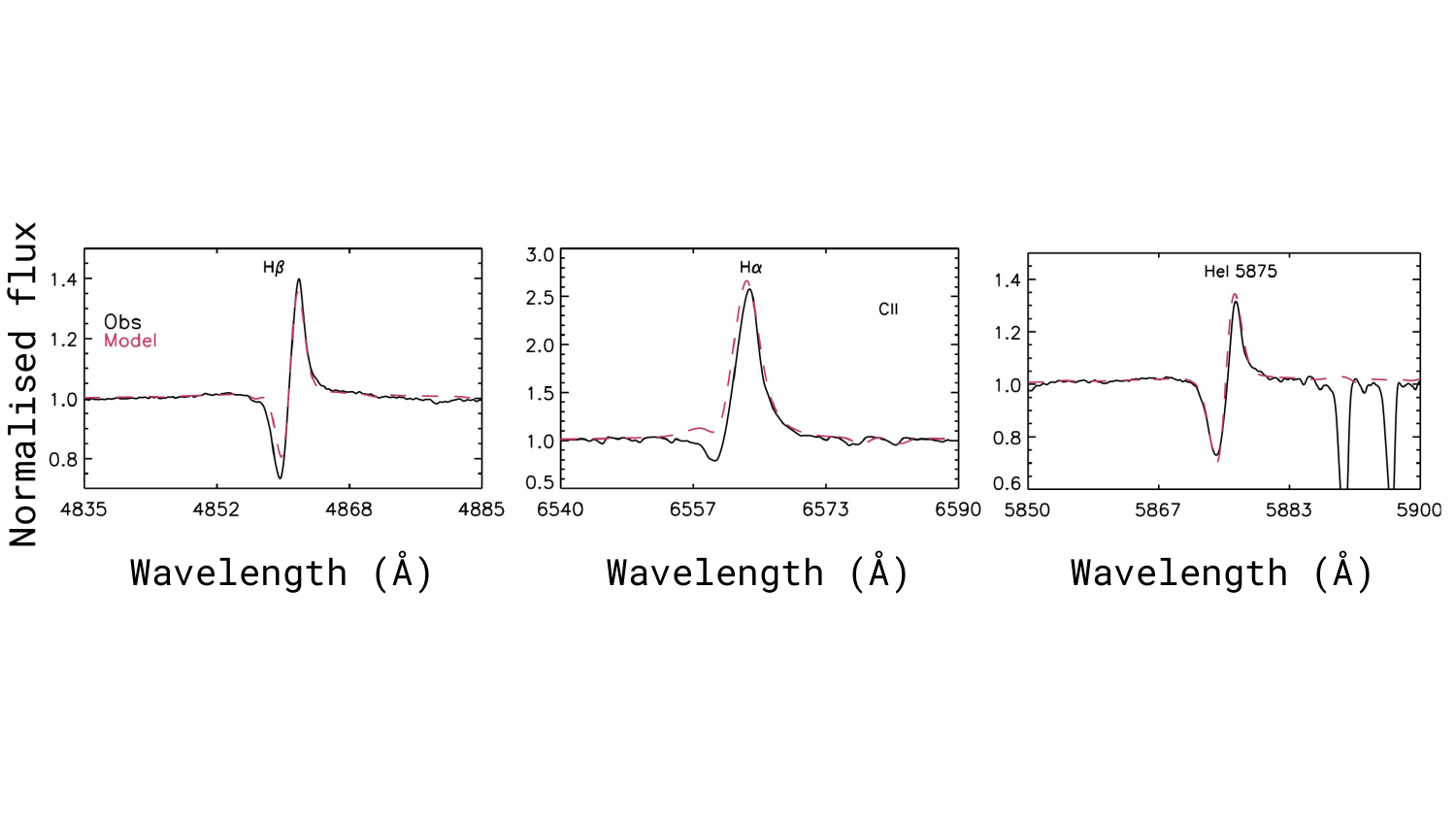}
\caption{Example of three P~Cygni line features in the H$\beta$, H$\alpha$ and He\,I~5875 lines of the B1.5\,Ia$^+$ BHG star $\zeta^1$~Sco (black solid lines). A model to the observed spectra is shown in with red dashed lines. A P~Cygni line profile in hydrogen Balmer and helium lines is a key distinguishing feature of BHGs with respect to their less luminous BSG counterparts. Adapted from~\citet{2012A&A...541A.145C}.}
\label{fig:ZetaSco}
\end{figure}

The presence of extreme winds are key features of hypergiant stars, which for these hot massive stars are line-driven winds (see the \textit{Stellar Winds} and \textit{Stellar Atmospheres} chapters for more details). 
Such extreme winds alter significantly the spectroscopic appearance of these stars and must be accounted for in stellar modelling. 
In this sense, BHGs are distinguished from BSGs via the presence of P~Cygni emission line profiles in the hydrogen Balmer series~\citep{1992A&AS...94..569L} and some helium lines.
Figure~\ref{fig:ZetaSco} displays examples of P~Cygni profiles for $\zeta^1$~Sco, which is one of the most luminous examples of a BHG in the Milky Way. 

The hypergiants can be roughly considered in three groups: the early-BHGs, which in the Milky Way have spectral types from B0 to B4; the late-BHGs, which have spectral types from B5 to B9; and  YHGs, which have spectral types A to G. 
This distinction is made partly based on evolutionary pathways, partly based on spectral morphology and partly based on stellar properties.
The early-B hypergiants are the post-main-sequence descendants of stars with initial masses roughly in the range 30--60\,M$_\odot$ that may subsequently evolve to become LBVs~\citep{2012A&A...541A.145C,clark2018Q,2014A&A...564A..30G}.
These stars display luminosities in the range $5.5 < \log$L/L$\odot < 6.5$ that extend up to the peak of the stellar luminosity function and appear close to the Eddington limit, above which they are potentially unstable to continuum-driven super-Eddington winds (see the \textit{Stellar Atmospheres} chapter for more details).
Late-B-type hypergiants are, in general, less luminous and, as their later spectral types suggest, have lower effective temperatures. 
Based on their observational properties, these stars are more likely of initial lower mass ($\sim$30\,M$_\odot$) than their earlier hypergiant counterparts and may be more closely related to LBV stars and YHGs in a redward evolutionary sequence from the RHG regime.

YHGs are clearly in an evolved state and have experienced significant mass-loss events~\citep{1998A&ARv...8..145D}.
They are often described as occupying the \textit{yellow void} of the HRD, which references the rarity of evolved stars in the luminosity and temperature range occupied by these stars.
Their atmospheres are highly extended and highly active, and such stars display distinct quiescent and eruptive phases, where their effective temperatures are significantly different.
While YHGs and YSGs have been associated with objects displaying post-RSG evolution~\citep{1998A&ARv...8..145D}, observational evidence does not universally support this, instead favouring redward evolution towards the RSG phase~\citep{2022MNRAS.511.4360K}.
In addition, the evolutionary connection with LBVs is strong, with both objects showing similar characteristics.
This evolutionary scheme is RSG/RHG $\rightarrow$ YHG $\rightarrow$ LBV.

The number of stars classified as BHG in the Milky Way is around 25.
\citet{2012A&A...541A.145C} compiled the known early and late BHGs, which totals 16 (8 early- and 8 late-BHGs). 
In addition to this are the seven BHGs, since discovered in the Quintuplet cluster~\citep{clark2018Q}, and the 5 within the isolated population distributed throughout the Galactic centre region~\citep{2021A&A...649A..43C}. 
Westerlund-1 cluster hosts 5 BHGs, although the cohort in the Quintuplet clusters are exclusively early-BHGs and those in Westerlund-1 are late-BHGs.
Despite their high intrinsic luminosities, YHGs can be difficult to identify and classify in the Milky Way~\citep[e.g.][]{2023ApJ...952..113B}. \citet{1998A&ARv...8..145D} reviews YHGs in the Local Group of galaxies and lists their properties, and while the paper notes that \textit{We did not strive for completeness}, the list that is presented in Table~2 arguably remains the best reference for a list of YHGs.
It is common in the literature to use the classification of YHG for stars at the peak of the YSG luminosity function~\citep[e.g.][]{2005A&A...435..239C,2023AJ....166...50H} rather than using the spectroscopic appearance of the H$\alpha$ line in emission as defined by~\citet{1971CoKit.554...35K} and advocated by~\citet{1998A&ARv...8..145D}.

\subsection{Blue and yellow supergiants}

BSGs are those with spectral types B0--B9. The YSGs have a less well-defined system, but here I consider YSGs those with spectral types between A0 and G9 and K-type supergiants are considered as RSGs. 
Such a distinction varies as a function of metallicity and, in the literature, often one sees more specific language, such as BA-type supergiants, AF-type supergiants, etc.
This slight ambiguity of nomenclature in the literature likely stems from the fact that the origins of the population of BSGs and YSGs are currently not fully understood. 

BSGs are the brightest stars in star-forming galaxies at visual wavelengths. 
Because of their large intrinsic brightness, BSGs and YSGs are identified and studied in Local Group galaxies and beyond. 
This, combined with their intrinsically young age -- which means they have not had time to move away from their birth environment -- makes these stars excellent probes of present-day stellar metallicities of galaxies~\citep{Kudritzki2008}.
This technique has been employed to great effect within and beyond the Local Group of galaxies, and BSGs have become one of the most important indicators of stellar metallicities in the study of the chemical evolution of galaxies.
\citet[][and references therein]{2017ApJ...847..112D} compiled abundance analysis results for various galaxies and determined the mass-metallicity relation of galaxies as probed by BSGs. 

One of the most important observations of BSGs is their apparent overabundance with respect to the evolutionary channels from which they originate.
The best examples of this are the BSG populations of Local Group galaxies such as the LMC and SMC. 
If we assume that such objects originate only from stars evolving rapidly towards the RSG and use a simple timing argument using stars on the main sequence as calibration, BSGs are a factor of 100 to 1000 times overabundant with respect to the predictions from evolutionary models.
Such overabundance remains insufficiently explained.
There are various observational tools to understand the origin of BSGs and YSGs, which have been used to establish evolutionary links.
Determining luminosities and effective temperatures to place these stars on the HRD is one of the most used tools to determine evolutionary connections and understand the nature of individual stars and stellar populations.
However, this alone often does not paint the full picture.
Abundance patterns on the surface of stars can be vital to determine evolutionary stage.
Enrichments of elements processed by the CNO cycle and other key elements at the stellar surface are used as a signature of post-RSG evolution~\citep[e.g.][]{1998A&ARv...8..145D, trundle2005,trundle2007}.
Comparing mass-loss rates of stellar populations is sometimes used to assign an evolutionary connection~\citep[e.g.][]{2016ApJ...825...50G}.
In addition, variability and pulsational properties can also provide vital clues to evolutionary phases~\citep{2020ApJ...902...24D}. 
The $\alpha$~Cygni variables are an example of this, which are stars that display high luminosities and short-timescale variability. Based on this, these objects are considered post-RSG objects, but this is still to be definitively demonstrated.
\citet{2024ApJ...967L..39B} explore the potential of astroseismology to uncover the origin of BSGs by grouping them in terms of the variability properties, which shows great potential.


In the single star stellar evolution paradigm, BSGs and YSGs are considered as either pre- or post-RSG objects. 
Stars that are yet to experience a RSG phase are `pre-RSG objects' and are currently 
\textit{i.} still on the main sequence fusing hydrogen to helium in their cores, 
\textit{ii.} core-helium burning objects on a slow trajectory redwards towards the RSG phase 
or \textit{iii.} in a rapid expansion phase having exhausted their core-hydrogen fuel. 
The contribution of each of these channels to the population of observed BSGs (and to some extent YSGs) is not well constrained.
The extent of the main sequence in stellar evolutionary models requires calibration from observations and extends into the BSG regime.
Post-RSG objects are more evolved stars that have spent at least some portion of their lives as a RSG and are currently experiencing some form of blue-loop evolution. 
Both pre- and post-RSG objects are expected to subsequently evolve to the RSG phase once again, although based on the progenitor of SN1987A, which is one of the closest and best studied supernova explosions -- which exploded as a B3\,Ia star -- this is clearly not always the case. 

In addition to the single star evolutionary pathways for BSGs, it is likely that a significant population of these stars are binary interaction products: 
stellar mergers can also produce BSGs. 
Stars within a close binary system can interact and merge to create a rejuvenated star.
Blue straggler stars are probably the most famous observational example of such an event, which were first discovered in old globular clusters as anomalously bright, young objects.
Given the high rate of occurrence of massive stars in binary systems, stellar mergers are predicted to be reasonably common.
Although which binary systems will result in a merger, the merger event and the outcomes of such an event are uncertain~\citep[see][for details on merger pathways]{2024A&A...682A.169H}, stellar merger models can explain the properties of individual BSGs~\citep{1990A&A...227L...9P} and lead to long-lived BSG phases~\citep{2024A&A...686A..45S}.
Moreover, it is likely that stellar mergers contribute significantly to observed BSG populations~\citep{2017MNRAS.469.4649M,2024ApJ...963L..42M}. 


In the YSG regime, while not completely ruled out theoretically, it remains highly unlikely that these stars are core-hydrogen burning objects.
Core-helium burning stars and those experiencing rapid expansion likely contribute to the observed populations.
Multiple studies have identified populations of YSGs that can potentially be explained as post-RSG objects~\citep{1991A&A...245..593H,2016ApJ...825...50G,2023AJ....166...50H}, which is typically based on comparing the observed mass-loss rates and luminosities with those of RSGs.
However, the origin of the majority of the YSG population remains uncertain.
Similarly, stellar merger products and mass transfer within binary systems generally predict fewer YSGs than BSGs~\citep{2024A&A...686A..45S}.

\section{Red supergiant and hypergiant stars}\label{rsgs}

\begin{figure}[t]
\centering
\includegraphics[width=.5\textwidth]{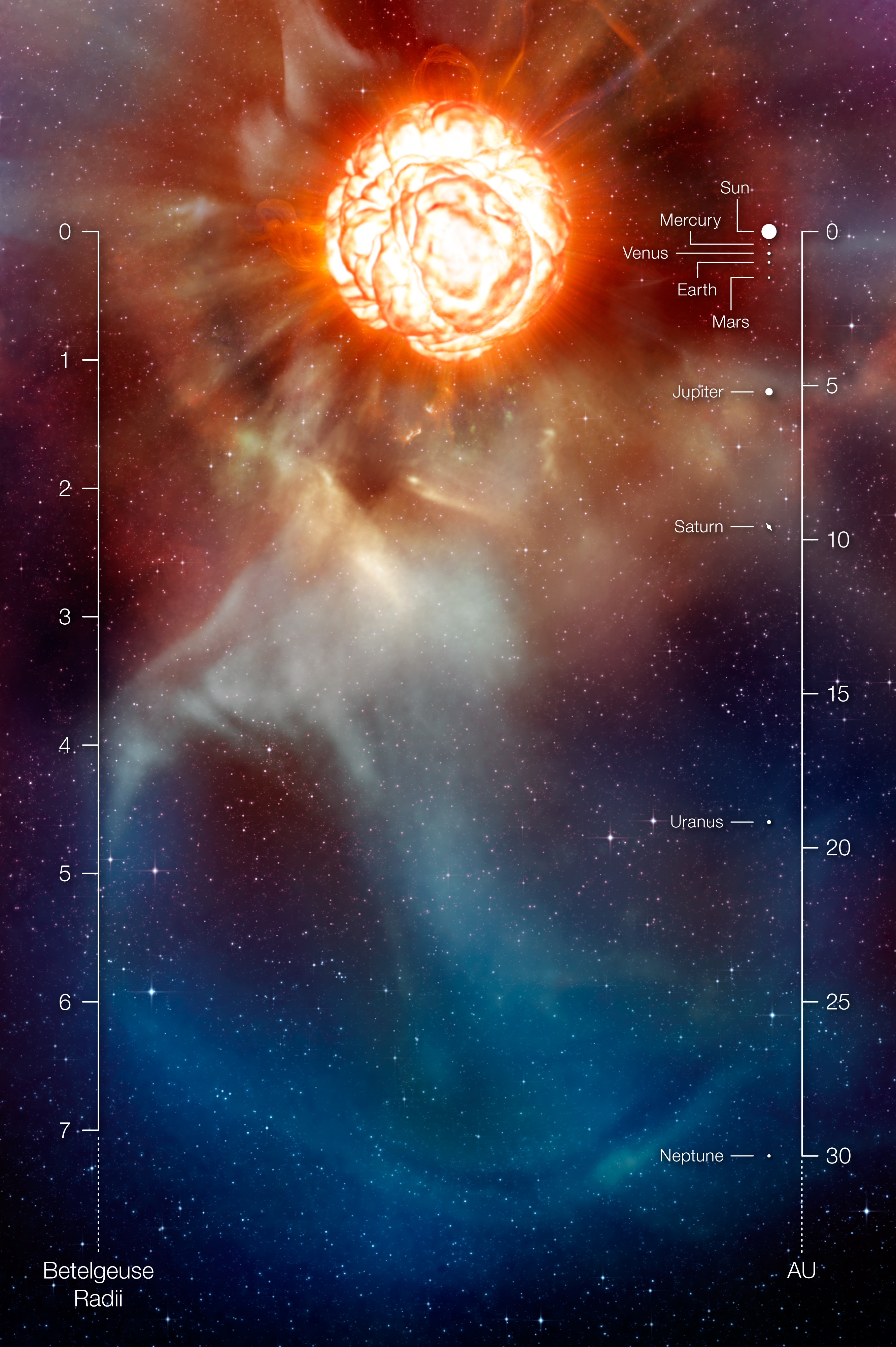}
\caption{Artist's impression of one of the most famous red supergiant stars, Betelgeuse and the scale of Betelgeuse shown against the Solar System. The distance from the centre of the star and the centre of the Solar system is shown in units of Betelgeuse radii and astronomical units on the left- and right-hand sides, respectively. The convective cells that dominate the surface of RSGs are illustrated. Material that has been expelled from the star via its intense wind is illustrated and demonstrates that the effects of mass loss extend many times past the radius of the star.
Credit: ESO/L. Cal\c{c}ada}
\label{fig:betelgeuse}
\end{figure}

Red supergiant stars are cool, luminous stars in the rough mass range $8 < M/M_\odot < 40$ that have evolved off the main sequence and are now burning helium in their cores. 
The majority of massive stars explode as supernova within the RSG phase, but for some it is a transition phase as the star returns to hotter temperatures. 
RSGs can be over 1000 times the radius of the Sun and up to hundreds of thousands of times more luminous. 
RSGs are important indicators of stellar evolution, and their properties and statistics are used to tune the physics that go into stellar evolutionary models.
Because of their high intrinsic luminosities, RSGs are the brightest stars in extragalactic stellar populations, and individual RSGs can be detected at large distances in external galaxies~\citep[e.g.][]{2017MNRAS.467..102C} and can be used to study elemental abundances in galaxies beyond the Local Group~\citep[e.g.][]{2017MNRAS.468..492P}.
The scale of these stars is often difficult to demonstrate effectively because, as with many astrophysical quantities, one must re-adjust ones reference points to comprehend stellar sizes.
One of the closest and most famous RSGs, Betelgeuse, is often compared with the scale of the Solar System. 
A press release from the European Southern Observatory is reproduced in Figure~\ref{fig:betelgeuse} to demonstrate that the radius of Betelgeuse, if placed in the Solar system, would extend to beyond the orbit of Jupiter.

The distinction between \textit{red supergiant} and \textit{red hypergiant} stars is one of luminosity: RHGs are more luminous.
RHG stars represent the tip of the luminosity function of the RSGs and are perhaps best thought of as the most massive and extreme examples of RSGs rather than a distinct population. 
Spectroscopically, the distinction between RSGs and RHGs is made based on the presence of the H$\alpha$ line in emission -- rather than absorption for a typical RSG -- which indicates the presence of a stellar wind more intense than a typical RSG. 
In practice, much of the literature on these two classes of objects is focused on RSGs.
RSGs are themselves a rarity in a stellar population (see Section~\ref{rsg_pops}) and RHGs are even rarer still, with only a handful of examples existing in entire populations of galaxies.

At the other end of the luminosity scale, \textit{red supergiant} stars are distinguished from \textit{red giant} stars based also on their luminosity; however, in this instance, this is because of an intrinsic difference in their evolutionary history.
After core-hydrogen fusion on the main sequence ceases, the evolution and eventual fate of a star with an initial mass lower than around 8\,M$_\odot$ is fundamentally different than for a star above this limit. 
More massive stars explode as supernova and lower-mass stars end their lives as an inert white dwarf star.
The distinction between the asymptotic giant branch stars (AGBs) and RSGs is intrinsically blurred, as such stars can have luminosities that rival RSGs~\citep[for an example, see the case of VX~Sagitarii;][]{2021A&A...646A..98T}.



\subsection{Stellar structure and evolution}

RSGs are in a late stage of stellar evolution. 
The lifetime of a RSG begins with core helium fusion via the triple $\alpha$ process with the reaction $3\alpha \rightarrow ^{12}C + \gamma$, and most of the time a stars spends as a RSG is within this phase.
As this nuclear burning process is highly temperature sensitive, a convective core is established surrounded by a hydrogen-fusing shell~\citep{1987A&A...182..243M}. 
At the end of core-helium fusion, the core of the star consists of inert carbon and oxygen surrounded by a helium-burning shell, which is in turn surrounded by a hydrogen-fusing shell.
The star then undergoes a period of core carbon fusion via the reaction $^{12}C (\alpha,\gamma) ^{16}O$.
This process repeats for $\alpha$-elements (oxygen, neon, magnesium, silicon, etc.), which are each more rapid than the previous, resulting in the famous \textit{onion structure} of RSGs, with an inert iron core unable to provide sufficient energy to support the star against gravitational collapse. 
Such an onion structure, containing shells of different fusion products, is an incredibly shortened lived period in the lifetime of a RSG.
Perhaps a more useful picture to have in mind to better comprehend the physical conditions in a typical RSG is that of core helium burning with a surrounding layer of hydrogen burning and a thick convective outer atmosphere. 
The ability to fuse carbon without the need for a degenerate core is typically used to distinguish between massive stars and intermediate mass stars. 

The outward appearance of RSGs is dominated by large convective cells~\citep{1975ApJ...195..137S}.
\citet{2024LRCA...10....2C} provide an in-depth review into the phenomenon of convection in RSGs. 
The convective cells of RSGs are on an enormous scale, such that several tens of cells can dominate the star. 
Such intrinsically active stellar atmospheres have a dramatic effect on the basic stellar properties of the star, such as their temperature and radii. 
Convective activity has been found to be highly correlated with temperature variations on the surface~\citep{2021A&A...650L..17K}.
RSGs also have well-studied pulsations~\citep{1969ApJ...156..541S}, of which several periodicities have been determined~\citep{2006MNRAS.372.1721K}, and have been linked to mass-loss events~\citep{2021A&A...650L..17K}.
While all of the pulsation modes are not well explained~\citep{2023MNRAS.526.2765S}, almost all RSGs show signatures of variability.

Mass-loss rates in RSGs are relatively poorly understood.
In the last 5 years, much work has gone into updating and revising earlier prescriptions of RSG mass-loss rates of~\citet{1988A&AS...72..259D} and \citet{1990A&A...231..134N}, which have been shown to be inadequate~\citep{2020MNRAS.492.5994B}. 
Multiple authors have recently tackled this problem from theoretical~\citep{2021A&A...646A.180K} and observational~\citep{2020MNRAS.492.5994B,2023A&A...676A..84Y,2024A&A...681A..17D} perspectives, the results of which are a significantly improved understanding of mass-loss rates as a function of luminosity and mass and their effects~\citep{2024arXiv241007335Z}.




Stellar evolution prior to the RSG phase occurs on the main sequence.
While it isn't clear if stars spend their entire core-helium burning lifetime as a RSG, it is expected that once core-hydrogen is exhausted on the main sequence, the star evolves on a thermal timescale to the RSG phase.
For stars in the rough mass range $8 < M/M_\odot < 15$, a blue loop is often predicted~\citep{2012A&A...537A.146E,Schootemeijer2019}.
Blue loops are transition phases where a RSG evolves to warmer temperatures and appears as a YSG. 
Their prevalence in RSG evolution is debated. 
Stars in the mass range  $8 < M/M_\odot < 23$ are expected to explode as supernova as a RSG. 
Once the onion structure described above is established, the star is unable to support itself under gravity and collapses.
Indeed, RSGs are the most common types of supernova progenitors and result in hydrogen-rich supernova with a plateau phase (Type II-P)~\citep[see ][for a review of supernova progenitors]{Smartt2009}. 
\citet{Smartt2009,2015PASA...32...16S} highlighted a tension between observed populations of RSGs in the Local Universe and the progenitors of SNII-P, which is termed \textit{the red supergiant problem} in the literature.
Effectively, there are too few supernova progenitors at high RSG masses (above $\sim$16--20\,M$_\odot$) to account for the theoretically predicted and observed RSGs in this mass range. 
RSGs have also been observed to end their lives without a supernova explosion.
Disappearing in 2009, N6946-BH1~\citep{2015MNRAS.450.3289G,2017MNRAS.468.4968A} is thought to have been observational proof of such a failed-supernova from a RSG.
However, further observation revealed that the interpretation of this event is not conclusive~\citep{2024ApJ...964..171B} and a stellar merger is likely to be able to explain all of the observations without the need to invoke a failed supernova.


Stars more massive than around 25\,M$_\odot$ are predicted to permanently evolve out of the RSG phase and end their lives in a hotter evolutionary stage. 
While apparently ubiquitous in stellar models, and supported by supernova progenitor observations, such evolution is not well understood.
In particular, the connection with YHGs and BHGs, as well as LBVs, is not well constrained.
There are rare examples of stars being observed to transition out of the RSG phase (e.g. WOH G64 in the LMC, Munoz-Sanchez et al. \textit{submitted}), which appears to have resulted in a B[e]SG, which are poorly understood, rare evolutionary phases that share many physical characteristics with LBVs~\citep{2019Galax...7...83K}.


\subsection{Observed properties and populations} \label{rsg_pops}
As is detailed in the \textit{Stellar Classification} chapter, the spectral appearance of RSGs is dominated at visual wavelengths by densely spaced molecular features from titanium oxide (TiO).
These stars' spectral types of typically late-K and M-types and are type Ia, Iab or Ib in luminosity class.
As their effective temperatures are $\sim$3000-4000\,K, their peak luminosity is around 800nm.
At these wavelengths, RSGs are the brightest stars in any star-forming galaxy.
The number of RSGs expected in a Milky Way-sized galaxy is on the order of 10\,000, but as a result of our position within the Milky Way, combined with confusion of the far more abundant lower-mass red giant stars, which appear similar in their photometric colours, the number of known RSGs in the Milky Way is less than 5\,\% of this number~\citep{Messineo_Brown_2019,Healy_2024}.
In external galaxies, this problem of confusion is significantly alleviated, because such galaxies are located out of the plane of the Milky Way, which enables much smaller relative uncertainties on distances. 
Such an advantageous location outside our own Galaxy is a significant advantage insofar as it allows the identification and characterisation of an almost-complete population of RSGs and RHGs.
This is important because, while in the Milky Way individual RSGs can be studied in great detail (e.g. Betelgeuse and Antares), often what we miss in our interpretation of individual systems is context in the wider population.
An example of this is perhaps the Great Dimming of Betelgeuse. 
This was an event where one of the brightest stars in the northern sky, Betelgeuse, was observed fainter than usual by more than one magnitude~\cite{2020ATel13512....1G}, which was visible even to the naked eye.
The origin of this event can likely be understood as an outflow during a pulsation cycle~\citep{2021A&A...650L..17K}.
Such a dramatic dimming was interpreted as an indicator of an immanent explosion~\citep{2023MNRAS.526.2765S} or evidence of a binary companion~\citep{2025ApJ...978...50M}, but what we lack in this context is how frequently such events occur in RSGs in general.


Two of the most well-studied populations of RSGs in the Local Universe are found in our two nearest-neighbour galaxies: the SMC and LMC.
In these galaxies, as a result of the recent advances from the Gaia mission, specifically the second and third data releases, which provided astrometric information for a significant fraction of the stellar population of these galaxies~\citep{2021A&A...649A...7G}, the known population of RSGs is near-complete~\citep{2020ApJ...900..118N,2021A&A...646A.141Y,2021ApJ...923..232R, 2021ApJ...922..177M,2023A&A...676A..84Y}.
A complete population as a function luminosity allows us to test population synthesis models, which provides insight into the underlying physics that govern stellar evolution. 
In addition, studies of a range of RSGs in galaxies at different metallicities show that the typical spectral appearance of RSGs, which are mostly M- and sometimes K-type supergiants at solar-like metallicities, transition to earlier spectral types at lower metallicities~\citep{2012AJ....144....2L, 2018MNRAS.476.3106T}.
In addition, spectral variability seems to be intrinsically larger at lower metallicities.

Similarly to BSGs, RSGs are key indicators of the chemical evolution of galaxies~\citep{2017ApJ...847..112D}.
Their relative youth means that they have not had time to travel significant distances from their birth sites, which means that such stars are probes of the elemental abundances\footnote{While \textit{chemical abundance} is a phrase that is much used in the literature, \textit{elemental abundance} is more accurate.} of their formation sites.
This technique has been employed to great success in the Local Group and beyond
and, making use of the fact that RSGs dominate the light from other stars~\citep{2014ApJ...787..142G,2016MNRAS.458.3968P}, integrated studies of young massive clusters containing multiple RSGs have been able to apply this technique to study elemental abundances to much greater distances~\citep{2015ApJ...812..160L}. 

Much like other evolutionary phases of massive stars, binarity is important in the lives of RSGs.
As RSGs are so intrinsically bright, identifying another source around such a star is often difficult; this is, of course, the principle that was used to great effect in the previous paragraph, but here it presents a problem. 
\citet{2024ARA&A..62...21M} summarise the recent progress in identifying and characterising RSG binary systems, ~\citet{2024IAUS..361..279P} provide a detailed comparison between these recent studies, and \citet{2024BSRSL..93..173P} summarise the properties of most well-studied RSGs in binary systems.
However, it is clear that our current picture is incomplete, because the most successful observational techniques in the hunt for RSG binary systems require the presence of an excess in the spectral energy distribution of RSGs that is attributed to a companion star~\citep{Neugent2018,2020ApJ...900..118N,2022MNRAS.513.5847P}.
These techniques are not sensitive to compact companions (black holes and neutron stars), nor do they provide any information on the orbital periods of individual systems. 

Samples of complete populations of RSGs remain important, even when they are not galaxy wide: star clusters are vital laboratories where the age of the stars in question can be considered effectively fixed. 
This has the advantage that effects such as evolution within the RSG phase can be determined. 
In studies where this has been done, the luminosity distribution of RSGs from stellar models that have a fixed age is not sufficient to reproduce the observed distributions~\citep{2025ApJ...981L..16W}.
What is required are models of significantly different ages.
The solution to this problem is merger products within binary systems, or \textit{red stragglers}~\citep{2019MNRAS.486..266B, 2019A&A...624A.128B}.
Red stragglers are the RSG analogues of blue straggler stars, which are binary merger products that appear to have ages that are too young based on the ages of the parent star cluster. 
As with single stars, once blue straggler stars exhaust the fuel in their cores, they evolve to the RSG phase, where they appear anomalously young in comparison with the RSG population that has not been affected by binary interactions. 
Estimates of the contribution of red stragglers to the overall population of RSGs are uncertain but significant: as many as 50\,\% of RSGs have been identified as red straggler stars in various young massive clusters~\citep{2019A&A...624A.128B,2020A&A...635A..29P}.
The uncertainty in these statistics arises because of the lack of agreement over how RSGs evolve within the RSG phase.
In some evolutionary models, significant vertical evolution is observed in the HRD (i.e. RSGs get more luminous over time), which would decrease the fraction of red stragglers. 

There are only a handful of stars classified as RHGs in the Local Universe.
Evidence of extreme mass loss is prevalent within such stars, and they represent the most luminous examples of RSGs.
In the Milky Way, the most famous example is probably VY~CMa, whose luminosity and mass loss make this star one of the most massive, evolved RSGs known. 
\citet{2019ApJ...874L..26H,2021AJ....161...98H} report on the various extreme mass-loss events that this star has experienced in the last several hundred years.
Estimates of the mass of such highly evolved RHGs are particularly uncertain~\citep{2020MNRAS.494L..53F}, but current mass determinations place VY~CMa in the 25--30\,M$_\odot$ range~\citep{2012A&A...540L..12W}.

\section{Low- and intermediate-mass red giants and supergiants} \label{giants}

The most common type of star when looking at the night sky is red giant stars.
This might seem paradoxical when one considers that the vast majority of the lifetime of a star is spent as a dwarf on the main sequence and that, consequently, the red giant phase represents only around 10\,\% of the lifetime of a star. 
The solution to this paradox is that stars like the Sun evolve almost vertically in the HRD after the main sequence, which means that red giant stars are many 100s or even 1000s of times brighter than they appeared on the main sequence.
Many of the stars described in this section have entire chapters dedicated to them within this encyclopedia and, therefore, where appropriate, we point the enthusiastic reader to their respective chapters.
In this chapter, the term \textit{red giant} is used as something of an umbrella term (different from the RGB), which describes stars at various evolutionary stages. 
This section describes low- and intermediate-mass giant and supergiant stars with spectral types A through M.
The distinction made here with respect to Section~\ref{ysgs} and~\ref{rsgs}, which also covers supergiants of similar spectral types, is that the giants and supergiants described here all end their lives by shedding their outer envelopes to become white dwarf stars.
There is a particular ambiguity here with the supergiant classification: Cepheid variables and AGB stars are supergiants that appear spectroscopically very similar to their more massive counterparts (i.e. YSGs and RSGs). 
Often, the only way to determine the evolutionary history of the star or stars in question is to determine stellar parameters derived from detailed spectroscopic observations and, even then, the picture is not always clear~\citep[for an example, see the case of VX~Sagitarii;][]{2021A&A...646A..98T}. 

\subsection{Stellar structure and evolution}
Core-hydrogen burning proceeds principally via the CNO cycle for stars greater than around 1.1\,M$_\odot$~\citep{2013sse..book.....K}.
Once low- and intermediate-mass stars exhaust their core-hydrogen burning fuel, the core is no longer supported against gravity and begins to collapse.
A hydrogen-fusing shell is established at the core boundary and, as a result of the mirror principle of shell burning, as the core contracts, the outer envelope expands. 
Observationally, this evolution results in the star being located at the base of the RGB.
As the core continues to contract, the luminosity of the star continues to increase and the stars reach the tip of the RGB, although the temperature of the star is largely unaffected. 
More massive stars proceed to helium burning smoothly, but the so-called helium flash is required for lower mass stars to kick-start helium burning.
Stars with Solar-like metallicity spend their core-helium burning lifetime in the so-called red clump. 
Lower metallicity stars, such as those observed in Globular clusters, burn helium on the horizontal branch.
As the core helium is depleted, the star evolves onto the AGB, where the luminosity is increased significantly. 
The main source of energy is now either the helium-fusing shell (early AGB stars) or, later, the hydrogen-fusing shell (thermally pulsing AGB stars; TP-AGB).
For more details on these evolutionary pathways, see the \textit{Evolution and Final Fates of Low- and Intermediate-mass Stars} chapter. 

\subsection{Observed populations and observational properties}
\subsubsection{Distance indicators}
Stars that can be used to measure distances and that can be observed at large distances hold a uniquely important role in our understanding of the structure and scale of the Universe. 
A degeneracy exists between distance and luminosity that in general cannot be broken observationally without detailed observations such as parallax measurements (see the \textit{Gaia} chapter).
Some physical processes within stars, however, occur at well-defined luminosities that do not depend strongly on metallicity.  See \citet{2018SSRv..214..113B} for an in-depth review of the three distance indicator techniques described here.

\textbf{Cepheid variables:} 
Cepheid variables are possibly the most famous example of such a star that is used for distance measurement. 
Classical Cepheid variables are stars that are rapidly crossing the so-called instability strip on the HRD. 
In this narrow region of parameter space, stars can become unstable to pulsations.
The period of pulsations of Cepheid variable stars has been shown to be tightly corrected with luminosity. 
By determining the period of pulsations, one can determine the intrinsic brightness of the star. 
Classical Cepheids are supergiant stars of spectral types F to K that exhibit variability in spectral types corresponding with pulsations. These stars are typically quoted in the literature as having initial masses in the range 2--20\,M$_\odot$, which depends strongly on metallicity. 
See the \textit{Pulsating Stars: Cepheids and RR Lyrae Stars} chapter for more details.

\textbf{RR~Lyrae stars:}
RR~Lyrae stars are another class of pulsating variable stars that have a well-determined period-luminosity relation~\citep[although not at visual wavelengths;][]{2004ApJS..154..633C}. Such stars are helium-burning stars on the horizontal branch and have the spectral appearance as A--F giants, which, similar to the cepheids, show variability. 
RR~Lyrae stars are significantly older than cepheids and they therefore trace different stellar populations and -- because of stellar redistribution in the galactic potential -- are more homogeneous in their distribution throughout the Milky Way.
See \citet{2004rrls.book.....S} for an in-depth review of these stars and the \textit{Pulsating Stars: Cepheids and RR Lyrae Stars} chapter for more details.

\textbf{Tip of the RGB:}
Stars evolving along the RGB reach maximum luminosity before the point of core helium ignition, which acts to drastically decrease their luminosities and begins their evolution on the horizontal branch.
The maximum luminosity reached by stars that experience this helium flash is not a strong function of initial mass and, because of this, in stellar clusters and galaxies there is an effective limit to the brightness that such stars achieve.
As this limit is directly tied to well-understood physical principles, if this feature can be observed in stellar populations, the distance can be well constrained.

\subsubsection{AGB stars}
The AGB phase occurs for low- and intermediate-mass stars that establish degenerate carbon-oxygen cores. 
This phase of evolution represents the final throws of the nuclear burning lifetime of a star and directly proceeds the shedding of the outer layers of the star towards the white dwarf phase of evolution. 
The AGB phase is uniquely important, as it is in this phase that nucleosynthesis of heavy \textit{s}-process elements takes place. 
This -- combined with the physical process that takes place in AGB stars to dredge nuclear processed material to the surface of the star and the strong mass loss that takes place during this phase of evolution -- means that AGB stars produce and distribute the products of nucleosynthesis throughout the Universe. 

The AGB phase is split into two main phases: the early-AGB phase and the thermal pulse phase (TP-phase). 
The early-AGB phase occurs after the exhaustion of the helium core burning where helium shell burning takes places. 
In this phase the hydrogen shell burning provides little energy and the helium-burning shell immediately surrounding the inert carbon-oxygen core adds material to the core. 
More massive AGB stars undergo a so-called second dredge-up event. 
The first dredge-up event takes places as stars exhaust their core-hydrogen fuel and become red giants. 
The second dredge-up follows the formation of the degenerate carbon-oxygen core and is a mechanism whereby nuclear material from the CNO cycle is raised to the surface. This process changes the surface abundances of AGB stars. 

As the AGB star evolves, the helium-burning shell progresses outwards from the core until it reaches a layer of the star almost devoid of helium.
This thin shell is thermally unstable and produces periodic pulses known as thermal pulses. 
The energy source during the TP-phase is a combination of helium shell and hydrogen shell burning. 
Each pulse penetrates deeper into the star and eventually reaches the region between the shell burning layers. 
When this occurs, a third dredge-up event happens that pollutes the envelope of the star with nuclear processed material from the region between the shells. 
The high mass-loss rates that are experienced during this phase mean that this phase of evolution is almost entirely responsible for the abundance of carbon that we observe today. 




Because of their intrinsic brightness, large populations of AGB stars are detected in galaxies in the Local Group.
AGB stars are K- and M-type stars with a luminosity class ranging from giants to supergiants. 
In this respect, such stars are easily confused with the more massive RSGs. Studies employ brightness cuts below/above which RSG/AGBs do not appear, but such criteria are only partially effective. 
In practice, even with stellar parameters determined from detailed spectroscopic observations, it can be difficult to tell these objects apart.
Typically, the combination of the appearance of the Li\,I 6708$\AA$ and Rb features in the spectrum is an AGB star signature.
This is because massive stars do not produce rubidium in large quantities. Likewise, lithium is easily destroyed in stellar atmospheres and this means that any significant detection of the lithium line is a sign that lithium is actively being produced in the star. 
This process is known to happen in AGB stars, but not in RSGs. 
However, this has been called into question by the recent detection of lithium in massive RSGs~\citep{2024IAUS..361..410N}. 
Near-IR wavelengths are particularly favourable for AGB stars, as the peak luminosity of these stars is around 1\,$\mu$m. 
TP-AGB stars have been shown recently to be an accurate distance indicator~\citep{2023ApJ...956...15L}.

Throughout the AGB phase, mass is lost via strong stellar winds and the AGB phase is ended as a result of mass loss. 
Towards the end of their lives, AGB stars can enter a so-called superwind phase that effectively strips off the outer layers of the star. Stars in this phase are sometimes given an OH/IR designation, which indicates that such stars have strong infrared excesses.
As the outer envelope is stripped away, the resulting carbon-oxygen core is observed as a white dwarf star. 

For more details on these stars, the \textit{AGB Stars} chapter and \citet{1983ARA&A..21..271I} provide in-depth reviews into AGB stars.


\section{Summary}\label{summary}
Giant, supergiant and hypergiant are luminosity classes in the Morgan \& Keenan classification system~\citep[][]{Morgan1973} that signify stars that are more luminous than the majority of stars of their spectral type. 
This morphological classification has physical meaning, as giants, supergiants and hypergiants, in general, represent stars that have ceased burning hydrogen to helium in their cores and are now in a late stage of evolution. 
This picture breaks down, however, for the OB stars, which, depending on their initial mass, evolve through the giant, supergiant and hypergiant luminosity classes all while fusing hydrogen to helium in their cores. 
Stars born with masses greater than around 8\,M$_\odot$ start their lives as OB-type stars and evolve to become giant stars and, subsequently, BSGs, YSGs and RSGs. 
Blue loops may allow RSGs to evolve back to the BSG or YSG phase before potentially returning to the RSG phase before exploding as a core-collapse supernova. 
More massive stars are born as giant, supergiant or even hypergiant stars, depending on their mass, and do not evolve to the RSG phase. 
Hypergiant stars have extreme luminosities and evidence for significant mass loss. These stars may form an evolutionary sequence for the most massive stars to reach the RSG phase; in this respect, a RHG expels its outer layers and evolves into a YHG, which may subsequently evolve bluer.

Low- and intermediate-mass stars evolve to the RGB as they exhaust their core-hydrogen supply, where they are observed as K- and M-type red giant stars, and evolve through the early-AGB to the TP-AGB phase before eventually shedding their outer envelopes to become white dwarf stars.
Red giants in different evolutionary stages, in particular Cepheids and RR~Lyrae stars, are vital distance indicators in the Local Group and are used as a critical anchor to measure the scale of the Universe.

\begin{ack}[Acknowledgments]

The author would like to thank Fabian Schneider, Ignacio Negueruela and Melanie Woodward for their effort, comments and input on this manuscript.
In addition, the author would like to thank the massive star group at the Centro de Astrobiologia for discussions and suggestions.  
L.R.P. acknowledges support by grants
PID2019-105552RB-C41 and PID2022-137779OB-C41 funded by
MCIN/AEI/10.13039/501100011033 by "ERDF A way of making
Europe". 
\end{ack}

\seealso{\citet{2017ars..book.....L}, Astrophysics of Red Supergiants, \citet{2013NewAR..57...14M} review massive star populations in the star-forming galaxies of the Local Group. \citet{2004agbs.book.....H}, Asymptotic Giant Branch Stars.}

\bibliographystyle{Harvard}
\bibliography{reference}

\end{document}